\documentclass[11pt]{article} 
\usepackage{amsmath}
\usepackage{amsthm}
\usepackage{epsfig}
\input{amssym.def}

\newtheorem{theorem}{Theorem}
\newtheorem{lemma}{Lemma}
\newtheorem{corollary}[theorem]{Corollary}

\newenvironment{singlespace}{\large\normalsize}{\par}
\def\noi{\noindent}

\def\parsec{\par\noindent}

\def\P{{\Bbb P}}
\def\E{{\bf E}\,}

\def\eps{\varepsilon}

\def\Var{{\bf Var}\,}
\def\Po{{\rm Po}}

\newcounter{thmenumerate}

\newenvironment{claim*}{\par\em}{\par}

\newcommand{\refT}[1]{Theorem~\ref{#1}}
\newcommand{\refC}[1]{Corollary~\ref{#1}}
\newcommand{\refL}[1]{Lemma~\ref{#1}}

\newcommand{\refS}[1]{Section~\ref{#1}}

\newcommand{\refand}[2]{\ref{#1} and~\ref{#2}}

\newcommand\set[1]{\ensuremath{\{#1\}}}

\newcommand\bigpar[1]{\bigl(#1\bigr)}
\newcommand\Bigpar[1]{\Bigl(#1\Bigr)}

\def\rompar(#1){\textup(#1\textup)}    % usage: \rompar(...)
\newcommand\xfrac[2]{#1/#2}

\newcommand\parfrac[2]{\Bigpar{\frac{#1}{#2}}}

\newcommand\ceil[1]{\lceil#1\rceil}
\newcommand\floor[1]{\lfloor#1\rfloor}

\newcommand\ntoo{\ensuremath{{n\to\infty}}}

\newcommand\iid{i.i.d.\spacefactor=1000}     %?????
\newcommand\ie{i.e.\spacefactor=1000}

\newcommand\cf{cf.\spacefactor=1000}

\newcommand\pto{\overset{\mathrm{p}}{\to}}
\newcommand\psim{\overset{\mathrm{p}}{\sim}}

\renewcommand\={:=}

\newcommand\Bi{\operatorname{Bi}}

\newcommand\ga{\alpha}

\newcommand\gd{\delta}

\newcommand\gam{\gamma}

\newcommand\kk{\kappa}
\newcommand\gl{\lambda}

\newcommand\bX{{\overline X}}

\newcommand\sumjk{\sum_{j=0}^k}
\newcommand\muj{\mu_j}
\newcommand\tF{\tilde F}
\newcommand{\tfx}[1]{\tF_{#1}}
\newcommand{\tfl}{\tfx{\gl}}
\newcommand{\tfxa}[1]{\tfx{#1}(\ga)}
\newcommand{\tfla}{\tfl(\ga)}
\newcommand{\kl}{k_\gl}

\newcommand\ddx[1]{\frac{d}{d#1}}
\newcommand\ddy[2]{\frac{d#1}{d#2}}
\newcommand\qx{^{1/2}}
\newcommand\qq{^{-1/2}}
\newcommand\qi{^{-1}}

\newcommand\afillup{$\ga$-fillup}
\newcommand\xii{\xi_1}

\newcommand\ones{N_1}
\renewcommand\Pr{P(r)}

\newcommand\whp{whp}

\begin{document}

\begin{center}
{\LARGE {\bf Partial Fillup and Search Time in LC Tries}}
\medskip
\parsec
\today
\end{center}
\par
\bigskip\bigskip
\begin{singlespace}
\noi
\begin{tabular}{l@{\hspace{1.25in}}l}
Svante Janson&Wojciech Szpankowski\footnotemark\\
Department of Mathematics&Department of Computer Science\\
Uppsala University, P.O. Box 480&Purdue University\\
SE-751 06 Uppsala&W. Lafayette, IN 47907\\
Sweden&U.S.A.\\
{\tt svante.janson@math.uu.se}&{\tt spa@cs.purdue.edu}
\end{tabular}
\footnotetext{The work of this author was supported in part by the NSF
Grants CCR-0208709, CCF-0513636, and DMS-0503742, 
NIH Grant R01 GM068959-01, and and AFOSR Grant FA8655-04-1-3074.}

\begin{abstract}
Andersson and Nilsson introduced in 1993 a {\it level-compressed trie}
(in short: LC trie) in which a full subtree of a node is compressed
to a single node of degree being the size of the subtree. 
Recent experimental results indicated a ``dramatic improvement'' 
when full subtrees are replaced by ``partially filled subtrees''. 
In this paper, we provide a theoretical justification of these experimental 
results showing, among others, a rather moderate improvement of the
search time over the 
original LC tries. For such an analysis, we assume that
$n$ strings are generated independently by a binary memoryless source
(a generalization to Markov sources is possible) with
$p$ denoting the probability of emitting a ``$1$'' (and $q=1-p$). We first
prove that the so called $\alpha$-fillup level $F_n(\alpha)$ 
(i.e., the largest level in a trie with $\alpha$ fraction of nodes 
present at this level) is concentrated on two values \whp{}
(with high probability); either $F_n(\alpha) =k_n $ or $F_n(\alpha) =k_n +1$ 
where
$
k_n=
\log_{ \frac{1}{\sqrt{pq} }} n
- \frac{ |\ln (p/q)|}{2 \ln^{3/2} (\xfrac{1}{\sqrt{pq}}) }
{ \Phi^{-1} ( \alpha )}
\sqrt{ \ln n } +  O (1)
$
is an integer and $\Phi(x)$ denotes 
the normal distribution function. This result
directly yields the typical depth (search time) $D_n(\alpha)$ in the
$\alpha$-LC tries with $p\neq1/2$, namely we show that \whp{}
$D_n(\alpha) \sim C_1 \log \log n$ where 
$C_1=1/|\log(1-h/\log(1/\sqrt{pq}))|$ and
$h=-p\log p -q \log q$ is the Shannon entropy rate. This should be compared 
with recently found typical depth in the original LC tries which is 
$C_2 \log \log n$  where 
$C_2=1/|\log(1-h/\log(1/\min\{p,1-p\}))|$.
In conclusion, we observe that $\alpha$ affects
only the lower term of the $\alpha$-fillup  level $F_n(\alpha)$, and
the search time in $\alpha$-LC tries is of the same order 
as in the original LC tries. 
\end{abstract}

\end{singlespace}

\newpage

\section{Introduction}

\begin{figure}
\centerline{\epsfig{file=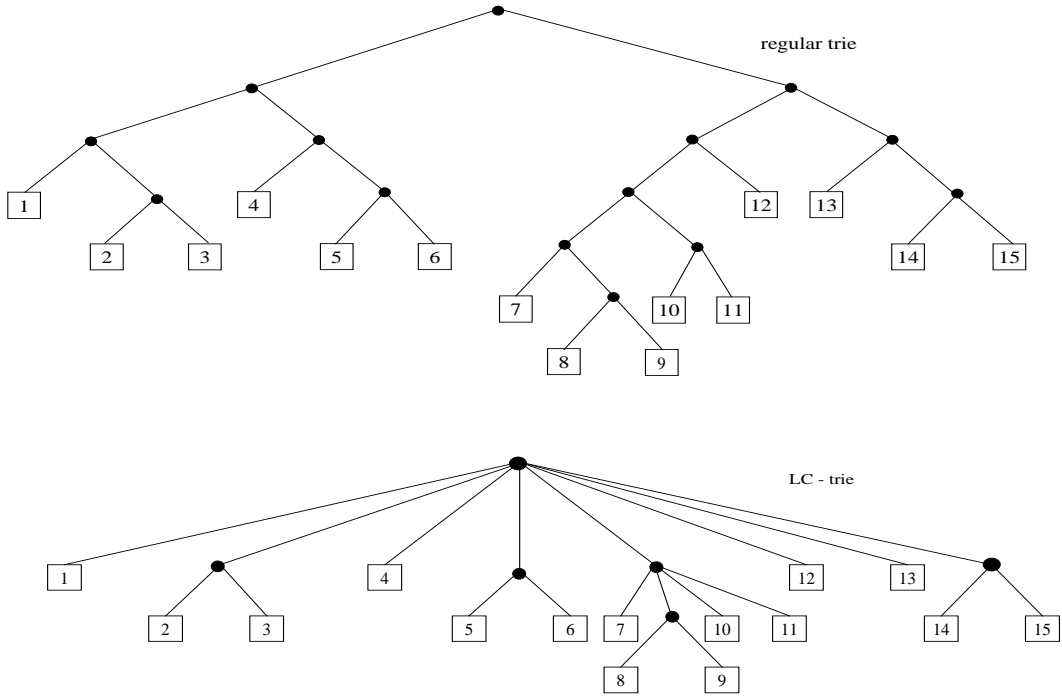,height=3.6in,width=5.5in}}
\caption{A trie and its associated full LC trie.}
\label{fig-trie}
\end{figure}

Tries and suffix trees are the most popular data structures on
words \cite{gusfield97}. 
A {\it trie} is a digital tree built over, say $n$, strings (the
reader is referred to \cite{knuth97,mahmoud,spa-book} for an in
depth discussion of digital trees.) A string is stored in an
external node of a trie and the path length to such a node is the
shortest prefix of the string that is not a prefix of any other
strings (\cf{} Figure~\ref{fig-trie}). 
Throughout, we assume a binary alphabet. 
Then each branching node in a trie is a binary node.
A special case of a trie structure is
a {\it suffix trie} (tree) which is a trie built over suffixes of
a {\it single} string.

Since 1960 tries were used in many
computer science applications such as searching and sorting,
dynamic hashing, conflict resolution algorithms, leader election
algorithms, IP addresses lookup, coding, polynomial factorization,
Lempel-Ziv compression schemes, and molecular biology. For
example, in the internet IP addresses lookup problem
\cite{nilsson,sv} one needs a fast algorithm that directs an
incoming packet with a given  IP address to its destination. As a
matter of fact, this is the {\it longest matching prefix} problem,
and standard tries are well suited for it.  However, the search
time is too large. If there are $n$ IP addresses in the database,
the search time is $O(\log n)$, and this is not acceptable. In
order to improve the search time, 
Andersson and Nilsson \cite{an93,nilsson} introduced 
a novel data structure called the {\it level compressed trie} 
or in short LC trie (\cf{} Figure~\ref{fig-trie}). 
In the LC trie we replace the root with a node of
degree equal to the size of the largest {\it full subtree}
emanating from the root (the depth of such a subtree is called the
{\it fillup level}). This is further carried on recursively
throughout the whole trie (cf.\ Figure~\ref{fig-trie}).  

Some recent experimental results reported in \cite{int99,nt02,nk99}
indicated a ``dramatic improvement'' in the search time when full subtrees
are replaced by ``partially fillup subtrees''. In this paper, we provide
a theoretical justification of these experimental results 
by considering $\alpha$-LC tries in which one replaces a subtree with
the last level only $\alpha$-filled by a node of degree equal to
the size of such a subtree (and we continue recursively). 
In order to understand theoretically the $\alpha$-LC trie behavior,
we study here the so called {\it $\alpha$-fillup}
level $F_n(\alpha)$ and the {\it typical depth} or the search time 
$D_n(\alpha)$. 
The $\alpha$-fillup level is the last level in a trie that is
$\alpha$-filled, \ie{}
filled up to a fraction at least $\alpha$ 
(e.g., in a binary trie level $k$ is $\alpha$-filled if it
contains $\alpha 2^k$ nodes).
The typical depth is the length of a path from the root to a randomly
selected external node; thus it represents the typical search time. 
In this paper we analyze the $\alpha$-fillup level and the typical depth in
an $\alpha$-LC trie
in a probabilistic framework when all strings are generated by a memoryless
source  
with $\P(1)=p$ and $\P(0)=q:=1-p$. 
Among other results, we prove that the $\alpha$-LC trie shows   
a rather moderate improvement over the original LC tries. We shall
quantify this statement below.

Tries were analyzed over the last thirty years for memoryless and
Markov sources (\cf{}  
\cite{devroye92,js91,ks05,knuth97,mahmoud,pittel85,pittel86,spa91,spa-book}).
Pittel \cite{pittel85,pittel86} found the typical value of the
fillup level $F_n$
(i.e., $\alpha=1$) in a trie built over $n$ strings generated by 
mixing sources;  for memoryless sources with high probability (\whp)
\begin{equation*}
F_n \psim \frac{\log n}{\log (1/p_{\min})} = \frac{\log n}{h_{-\infty}}   
\end{equation*} 
where $p_{\min}=\min\{p,1-p\}$ is the smallest probability of 
generating a symbol and
$h_{-\infty} = \log (1/p_{\min})$ is the R\'enyi entropy of 
infinite order (\cf{} \cite{spa-book}). 
We let $\log:=\log_2$. 
In the above, we write $F_n \psim a_n$ to denote
$F_n/a_n \to 1$ in probability, that is, for any $\eps>0$ we have
$\P((1-\eps) a_n \leq F_n \leq (1+\eps) a_n) \to 1$ as $n\to \infty$.

This was further extended by Devroye \cite{devroye92}, and
Knessl and Szpankowski \cite{ks05} who, among other results,  proved that
the fillup level $F_n$ is concentrated on two points $k_n$ and $k_n+1$,
where $k_n$ is an integer 
%the closest integer less than
\begin{equation}
\label{efn}
\frac{1}{\log p_{\min}^{-1}}\left
(\log n -\log \log \log n\right)+O(1)
\end{equation}
for $p\neq1/2$. 
The depth in regular tries was analyzed by many authors who
proved that \whp{} the   
depth is about $(1/h) \log n$ (where
$h=-p \log p - (1-p) \log (1-p)$ is the Shannon entropy rate 
of the source) and that it is normally distributed 
when $p\neq1/2$ \cite{pittel86,spa-book}. 

The original LC tries were analyzed by Andersson and Nilsson \cite{an93}
for unbiased memoryless source and by Devroye \cite{devroye01} 
for memoryless sources (\cf{} also \cite{reznik02,reznik05}). The 
typical depth (search time) for regular LC tries was only
studied recently by Devroye and Szpankowski \cite{ds05} who
proved that for memoryless sources with $p\neq1/2$  
\begin{equation}
\label{edn}
D_n \psim 
\frac{ \log \log n}{-\log \left( 1 - \xfrac{ h }{ h_{-\infty} } \right)}.
\end{equation}

In this paper we shall prove some rather surprising results. First of all, for
$0< \alpha<1$ we show that the $\alpha$-fillup level $F_n(\alpha)$ is
\whp{} equal either to $k_n$ or $k_n+1$ where
\begin{equation}
\label{efa}
k_n=
\log_{ \frac{1}{\sqrt{pq} }} n
- \frac{ |\ln (p/q)|}{2 \ln^{3/2} (\xfrac{1}{\sqrt{pq}}) }
{ \Phi^{-1} ( \alpha )}
\sqrt{ \ln n } +  O (1).
\end{equation}
As a consequence, we find 
that if $p\neq1/2$, 
the depth $D_n(\alpha)$ of the $\alpha$-LC
is
for large $n$
typically about
\begin{equation*}
%\label{eda}
\frac{\log \log n}{-\log\left(1-\xfrac{h}{\log (1/\sqrt{pq})}\right)}.
\end{equation*}

The (full) $1$-fillup level $F_n$ shown in \eqref{efn} should be compared to 
the $\alpha$-fillup level $F_n(\alpha)$ presented in \eqref{efa}. Observe that
the leading term of $F_n(\alpha)$ is {\it not} the same as the
leading term of $F_n$ when $p\neq1/2$. 
Furthermore, $\alpha$ contributes only to the
second term asymptotics. When comparing the typical depths $D_n$ and
$D_n(\alpha)$ we conclude that both grow like $\log  \log n$
with two constants that do not differ by much. This comparison led us
to a statement in the abstract that the improvement of 
$\alpha$-LC tries over the regular LC tries is rather moderate.
We may add that for relatively slowly growing functions such as $\log \log n$
the constants in front of them do matter (even for large values of $n$)
and perhaps this led the authors of \cite{int99,nk99,nt02} to 
their statements.  

The paper is organized as follows. In the next section we present our 
main results which are proved in the next two sections. We first
consider a poissonized version of the problem for which we
establish our findings. Then we show how to depoissonize
our results  completing our proof.

\section{Main Results}

Consider tries created by inserting $n$ 
random strings of 0 and 1.
We will always assume that the strings are (potentially) infinite and
that the bits in the strings are independent random bits,         
with $\P(1)=p$ and thus $\P(0)=q:=1-p$; moreover we assume that
different strings are independent. 

We let 
$X_k$ := \#\set{\text{internal nodes filled at level }k}
and
$\bX_k := X_k/2^k$, \ie{} the proportion of nodes filled at level $k$.
Note that $X_k$ may both increase and decrease as $k$ grows, while
$$1\ge\bX_k\ge\bX_{k+1}\ge0.$$ 

Recall that the fillup level of the trie is defined as the last full
level, i.e.\ $\max\set{k:\bX_k=1}$, while 
the height is the last level with any nodes at all, i.e.\
$\max\set{k:\bX_k>0}$.
Similarly, if $0<\ga\le1$, 
the \afillup\ level $F_n(\alpha)$  
is the last level where at least a proportion
$\ga$ of the nodes are filled, i.e.\
$$
F_n(\alpha)=\max\set{k:\bX_k\ge\ga}. 
$$

We will in this paper study the 
\afillup\ level for a given $\ga$ with $0<\ga<1$ and 
a given $p$ with $0<p<1$.

We have the following result, where \whp{} means
with probability tending to 1 as \ntoo, and
$\Phi$ denotes the normal distribution function.
\refT{T1} is proved in \refS{Sdepo}, after first considering a
Poissonized version in \refS{Spo}.

\begin{theorem}\label{T1}
Let $\ga$ and $p$ be fixed with $0<\ga<1$ and $0<p<1$, and 
let $F_n(\ga)$ be the \afillup\ level for the trie formed by $n$
random strings as above.
Then, for each $n$ there is an integer 
\begin{equation*}
k_n= 
\log_{ \frac{1}{\sqrt{pq} }} n 
- \frac{ |\ln (p/q)|}{2 \ln^{3/2} (\xfrac{1}{\sqrt{pq}}) }
{ \Phi^{-1} ( \alpha )}
\sqrt{ \ln n } +  O (1)
\end{equation*}
such that \whp{}
$F_n(\ga)=k_n$ or $k_n+1$. 
Moreover, $\E\bX_{k_n}=\ga+O(1/\sqrt{\log n})$ for $p \neq 1/2$.
\end{theorem}

Thus the $\alpha$-fillup level
$F_n(\ga)$ is concentrated on at most two values; 
as in many
similar situations (\cf{} \cite{devroye92,ks05,pittel85,spa-book}), 
it is easily
seen from the proof that in fact for most $n$ it is concentrated on a
single value $k_n$, but there are transitional regimes, close to the
values of $n$ where $k_n$ changes,  where $F_n(\ga)$ takes two values
with comparable probabilities.

Note that when $p=1/2$, the second term on the right hand side
disappears, and thus simply $k_n=\log n+O(1)$; in particular, two different
values of $\ga\in(0,1)$ have their corresponding $k_n$ differing by
$O(1)$ only. When $p\neq1/2$, changing $\ga$ means shifting $k_n$ by
$\Theta(\log^{1/2}n)$. 
By \refT{T1}, \whp{} $F_n(\ga)$ is shifted by the same amounts.

%In both cases, the value of $\ga$ thus affects only lower order terms,
%while the leading term is independent of $\ga$. 
To the first order, we thus have the following simple result.

\begin{corollary}\label{CT1}
For any fixed $\ga$ and $p$ with $0<\ga<1$ and $0<p<1$, 
\begin{equation*}
F_n(\ga)= 
\log_{ \frac{1}{\sqrt{pq} }} n + O_p(\sqrt{ \ln  n });
\end{equation*}
in particular,
$F_n(\ga)/ \log_{ \xfrac{1}{\sqrt{pq}}} n\pto1$
as \ntoo.
\end{corollary}

Surprisingly enough, the leading terms of the fillup level  
for $\alpha=1$ and
$\alpha<1$ are quantitatively different for $p\neq1/2$. 
It is well known, as explained in
the introduction, that the regular fillup level
$F_n$ is concentrated on two points around 
%$\frac{1}{h_{\infty}} \log n$ (where, we recall $h_\infty=1/\log p_{\min}$, 
$\log n/\log(1/p_{\min})$, 
while  the partial fillup level $F_n(\alpha)$ concentrates around 
$k_n\sim\log n/\log(1/\sqrt{pq})$.
Secondly, the leading term of $F_n(\alpha)$ does not depend on $\alpha$
and the second term is proportional to $\sqrt{\log n}$,
while for the regular fillup level $F_n$
the second term is of order $\log \log \log n$.

Theorem~\ref{T1} yields several consequences for the behavior of 
$\alpha$-LC tries. In particular, it implies the typical behavior
of the depth, that is, the search time. Below we formulate 
our main second result concerning the depth for $\alpha$-LC tries 
delaying the proof to \refS{Slc}; \cf{} \eqref{edn} and \cite{ds05,reznik05}
for LC tries.

\begin{theorem}
\label{T3}
For any fixed $0<\alpha<1$ and $p\neq 1/2$ we have
\begin{equation}
\label{eT3}
D_n(\alpha) \psim
  \frac{\log \log n}{-\log\left(1-\frac{h}{\log (1/\sqrt{pq})}\right)}
\end{equation}
as $n\to \infty$ where $h=-p\log p -(1-p) \log (1-p)$ is the entropy rate of
the source.
\end{theorem}

As a direct consequence of Theorem~\ref{T3} 
we can numerically quantify experimental results recently
reported in \cite{nk99} where a ``dramatic improvement'' 
in the search time of $\alpha$-LC tries over the regular LC tries 
was observed. 
In a regular LC trie the search time is $O(\log \log n)$ with the constant
in front of $\log \log n$ being $1/\log(1-h/\log(1/p_{\min}))^{-1}$
\cite{ds05}. For 
$\alpha$-LC tries this constant decreases to 
$1/\log(1-h/\log(1/\sqrt{pq}))^{-1}$. 
While it is hardly a ``dramatic improvement'', the fact that we deal
with a slowly growing leading term $\log \log n$, may indeed
lead to experimentally observed significant changes in the search time.

\section{Poissonization}\label{Spo}

In this section
we consider a Poissonized version of the problem, where there are
$\Po( \lambda )$ strings inserted in the trie. 
We let $\tfla$ denote the \afillup\ level of this trie.

\begin{theorem}\label{T2}
Let $\ga$ and $p$ be fixed with $0<\ga<1$ and $0<p<1$, and 
let $\tfla$ be the \afillup\ level for the trie formed by $\Po(\gl)$
random strings as above.
Then, for each $\gl>0$ there is an integer 
\begin{equation}\label{t2}
k_\gl= 
\log_{ \frac{1}{\sqrt{pq} }} \gl
- \frac{ |\ln (p/q)|}{2 \ln^{3/2} (\xfrac{1}{\sqrt{pq}}) }
{ \Phi^{-1} ( \alpha )}
\sqrt{ \ln \gl } +  O (1)
\end{equation}
such that \whp{} (as $\gl\to\infty$)
$\tfla=k_\gl$ or $k_\gl+1$.
\end{theorem}

We shall prove Theorem \ref{T2} through a series of lemmas.
Observe first that a node at level $k$ can be labeled by a binary string of
length $k$, and that the node is filled if and only if at least two of
the inserted strings begin with this label.
For $r\in\set{0,1}^k$,
let $\ones(r)$ be the number of
ones in $r$, 
and let $\Pr = p^{\ones(r)} q^{k - \ones(r)}$
be the probability that a random string begins with $r$.
Then, 
in the Poissonized version, the number of inserted strings beginning with
$r\in\set{0,1}^k$ has a Poisson distribution 
$\Po({\gl\Pr})$, 
and these numbers are
independent for different strings $r$ of the same length.  
Consequently, 
\begin{equation}
  \label{erika}
X_k =  \sum_{ r\in \set{0,1}^k } I_{r}
\end{equation}
where $I_{r}$ are independent indicators with
\begin{equation}
  \label{emma}
\P(I_{r} = 1) = \P(\Po(\gl\Pr)\ge2)=1 - (1+\gl\Pr)e^{-\gl\Pr}.
\end{equation}

Hence,
\begin{align*}
%\sigma_k^2:&=
\Var (X_k ) =  \sum_{r\in\set{0,1}^k} P(I_r = 1)\bigpar { 1 - P (I_r = 1 )} 
< 2^k
\end{align*}
so $\Var(\bX_k)<2^{-k}$ and, by Chebyshev's inequality,
\begin{equation}
\label{bx}  
\P(|\bX_k-\E\bX_k|>2^{-k/3}) \to0.
\end{equation}
Consequently, $\bX_k$ is sharply concentrated, and it is enough to
study its expectation. (It is straightforward to calculate $\Var(X_k)$
more precisely, and to obtain a normal limit theorem for $X_k$, but we
do not need that.)

Assume first $p>1/2$.
\begin{lemma}
  \label{L1}
If\/ $p>1/2$ and
\begin{equation}\label{l1}
k= 
\log_{ \frac{1}{\sqrt{pq} }} \gl
- \frac{ \ln (p/q)}{2 \ln^{3/2} (\xfrac{1}{\sqrt{pq}}) }
{ \Phi^{-1} ( \alpha )}
\sqrt{ \ln \gl } +  O (1),
\end{equation}
then $\E\bX_k=\ga+O(k\qq)$.
\end{lemma}
\begin{proof}
Let $\rho=p/q>1$
and define $\gamma$ by $\gl p^{\gamma} q^{k-\gamma} =1$, i.e., 
$$
\rho^\gam=\left( \frac{p}{q} \right)^\gamma = \lambda^{-1} q^{-k},
$$
which leads to
\begin{equation}
  \label{b}
\gamma =  \frac{ k \ln(1/q) - \ln \lambda}{\ln(p/q)}.
\end{equation}

Let $\muj=\gl p^jq^{k-j}=\rho^{j-\gam}$.
By \eqref{erika} and \eqref{emma}, 
\begin{equation}
  \label{magnus}
\E\bX_k = 2^{-k}\sumjk \binom kj \P({\Po(\muj)\ge2}).
\end{equation}
If $j<\gam$, then $\muj<1$ and 
$$
\P({\Po(\muj)\ge2})< \muj^2<\muj.
$$
If $j\ge\gam$, then $\muj\ge1$ and 
$$
1-\P({\Po(\muj)\ge2})=(1+\muj)e^{-\muj}
\le 2\muj e^{-\muj}<4\muj\qi.
$$
Hence \eqref{magnus} yields, 
using $\binom kj \le \binom k{\floor{k/2}}=O(2^k k\qq)$,
\begin{equation}
  \begin{split}
\E\bX_k    
&= 
2^{-k}\sum_{j<\gam}\binom kj O(\muj)
+2^{-k}\sum_{j\ge\gam}\binom kj (1-O(\muj\qi))
\\
&= 
2^{-k}\sum_{j\ge\gam}\binom kj
+2^{-k}\sumjk\binom kj O(\rho^{-|j-\gam|})
\\&
= 
\P\bigpar{\Bi(k,1/2)\ge\gam}
+O(k\qq).
  \end{split}
  \label{david}
\end{equation}
By the Berry--Esseen theorem \cite[Theorem XVI.5.1]{FellerII},
\begin{equation}
  \label{jesper}
\P(\Bi(k,1/2)\ge\gam)
=1-\Phi\parfrac{\gam-k/2}{\sqrt{k/4}}
+O(k\qq).
\end{equation}

By \eqref{b} and the assumption \eqref{l1},
\begin{align}
%\begin{split}
\gamma - \frac{k}{2} 
&=
\frac{1}{\ln (p/q)} 
\left(k  \ln \frac{1}{q} - \ln  \lambda -\frac{k}{2}\ln \frac{p}{q}
\right)
\notag\\
&= \frac{1}{ \ln (p/q) }
\left( k  \ln \frac{1}{\sqrt{pq} } - \ln \lambda \right) 
\notag\\
&=
\frac{\ln (\xfrac{1}{\sqrt{pq}})}{\ln (p/q)} 
\left( k - \log_{\xfrac{1}{\sqrt{pq}} } \lambda \right)
\label{manne}\\
&=
-\tfrac12(\ln (\xfrac{1}{\sqrt{pq}}))\qq  \Phi\qi(\ga)
\sqrt{\ln\gl}
+O(1)
\notag\\
&=
- \tfrac12 \Phi\qi(\ga)k^{1/2}+O(1).
\notag
%\end{split}
\end{align}
This finally implies
\begin{equation*}
1-\Phi\parfrac{\gam-k/2}{\sqrt{k/4}}
=
1-\Phi(-\Phi\qi(\ga))+O(k\qq)
=
\ga+O(k\qq),
\end{equation*}
and the lemma follows by
\eqref{david} and \eqref{jesper}.
\end{proof}

\begin{lemma}
  \label{L2}
Fix $p>1/2$.
For every $A>0$, there exists $c>0$ such that if 
$|k-\log_{1/\sqrt{pq}}\gl|\le A k^{1/2}$, then
$\E \bX_k - \E\bX_{k+1} >c k\qq$.
\end{lemma}

\begin{proof}
  A string $r\in\set{0,1}^k$ has two extensions $r0$ and $r1$ in
$\set{0,1}^{k+1}$.
Clearly, $I_{r0}, I_{r1} \le I_{r}$, and if there are exactly 2 (or 3) 
of the inserted strings beginning with $r$, then
$I_{r0}+I_{r1}\le1<2I_r$.
Hence
\begin{equation}
  \label{sofie}
\E(2 X_k-X_{k+1}) 
= \sum_{r\in\set{0,1}^k} \E(2I_r-I_{r0}-I_{r1})
\ge  \sum_{r\in\set{0,1}^k} \P\bigpar{\Po(\gl\Pr)=2}.
\end{equation}
Let $\rho$ and $\gam$ be as in the proof of \refL{L1}, and let
$j=\ceil\gam$. Then $\muj=\rho^{j-\gam}\in[1,\rho]$ and thus
$\P(\Po(\muj)=2)\ge\frac12 e^{-\rho}$.
Moreover, by \eqref{manne} and the assumption,
\begin{equation*}
  |j-k/2|
\le
\frac{\ln (\xfrac{1}{\sqrt{pq}})}{\ln (p/q)} A k\qx +1 = O(k\qx).
\end{equation*}
Thus, if $k$ is large enough, we have by
the standard normal approximation of the binomial probabilities
(which follows easily from Stirling's formula, as found already by de
Moivre \cite{deMoivre})
\begin{equation*}
  2^{-k}\binom kj
= \frac{1+o(1)}{\sqrt{2\pi k/4}} e^{-2(j-k/2)^2/k}
\ge c_1 k\qq
\end{equation*}
for some $c_1>0$.
Hence, by \eqref{sofie},
\begin{equation*}
\E\bX_k-\E\bX_{k+1}
= 2^{-k-1} \E(2X_k-X_{k+1})
\ge
2^{-k-1} \binom kj \P(\Po(\muj)=2)\ge\frac{c_1 e^{-\rho}}4 k\qq
\end{equation*}
as needed.
\vskip-\baselineskip
\end{proof}

Now assume $p>1/2$.
Starting with any $k$ as in \eqref{l1}, we can by Lemmas
\refand{L1}{L2} shift $k$ up or down $O(1)$ steps and find $\kl$ as in
\eqref{t2} such that, for a suitable $c>0$,
$\E\bX_{\kl} \ge \ga+\tfrac12 c \kl\qq 
>\E\bX_{\kl+1}$ and 
$\E\bX_{\kl+2} \le \E\bX_{\kl+1}- c \kl\qq 
< \ga-\tfrac12 c \kl\qq $.
It follows by \eqref{bx} that \whp{} 
$\bX_{\kl}\ge\ga$ and 
$\bX_{\kl+2}<\ga$, and hence 
$\tfla=\kl$ or $\kl+1$.

This proves \refT{T2} in the case $p>1/2$.
The case $p<1/2$ follows by symmetry, interchanging $p$ and $q$.

In the remaining case $p=1/2$, all $\Pr=2^{-k}$ are equal. Thus, by
\eqref{erika} and \eqref{emma}, 
\begin{equation}
  \label{axel}
\E\bX_k = \P(\Po(\gl2^{-k})\ge2).
\end{equation}
Given $\ga\in(0,1)$, there is a $\mu>0$ such that 
$\P(\Po(\mu)\ge2)=\ga$. We take $\kl=\floor{\log(\gl/\mu)-1/2}$.
Then, 
$\gl2^{-\kl}\ge 2\qx\mu$ and thus 
$\E\bX_{\kl} \ge \ga_+$ for some $\ga_+>\ga$.
Similarly, $\E\bX_{\kl+2} \le \ga_-$ for some $\ga_-<\ga$, and the
result follows in this case too.
%\end{proof}

\section{Depoissonization}\label{Sdepo}

To complete the proof of Theorem~\ref{T1} we must depoissonize the
results obtained in Theorem~\ref{T2}, which we do in this section.

\begin{proof}[Proof of \refT{T1}]
Given an integer $n$, let $k_n$ be as in the proof of \refT{T2} with
$\gl=n$, and let $\gl_\pm=n\pm n^{2/3}$.
Then $\P(\Po(\gl_-)\le n))\to1$ and $\P(\Po(\gl_+)\ge n))\to1$
as \ntoo.  By monotonicity, we thus have 
\whp{} $\tfxa{\gl_-}\le F_n(\ga) \le \tfxa{\gl_+}$, and by \refT{T2} it
remains only to show that we can take $k_{\gl_-}=k_{\gl_+}=k_n$.

Let us now write $X_k(\gl)$ and $\bX_k(\gl)$, since we are working
with several $\gl$.

\begin{lemma}
  \label{L3}
Assume $p\neq1/2$. Then, for every $k$,
\begin{equation*}
  \ddx\gl \E\bX_k(\gl) = O(\gl\qi k\qq).
\end{equation*}
\end{lemma}

\begin{proof}
  We have 
\begin{equation*}    
\ddx\mu\P(\Po(\mu)\ge2)=
\ddx\mu((1-(1+\mu)e^{-\mu})
=\mu e^{-\mu}
\end{equation*}
and thus, by \eqref{magnus} and the argument in \eqref{david},
\begin{equation*}
  \begin{split}
\ddx\gl\E\bX_k(\gl)
&=
2^{-k}\sumjk\binom kj \muj e^{-\muj}\ddy{\muj}\gl
\\&
=\gl\qi 2^{-k}\sumjk\binom kj \muj^2e^{-\muj}
=O\Bigpar{\gl\qi \sumjk 2^{-k}\binom kj \min(\muj,\muj\qi)}
\\&
=O(\gl\qi k\qq)
  \end{split}
\end{equation*}
which completes the proof.
\vskip-\baselineskip
\end{proof}

By \refL{L3},
$|\E\bX_k(\gl_\pm)-\E\bX_k(n)|
=O(n^{-1/3}k\qq)
=o(k\qq)
$.
Hence, by the proof of \refT{T2}, for large $n$, 
$\E\bX_{k_n}(\gl_\pm)\ge\ga+\tfrac13ck_n\qq$
and
$\E\bX_{k_n+2}(\gl_\pm)<\ga-\tfrac13ck_n\qq$, and thus \whp{}
$\tfxa{\gl_\pm} = k_n$ or $k_n+1$.
Moreover, the estimate $\E\bX_{k_n}=\ga+O(1/\sqrt{\log n})$ follows 
easily from the similar estimate for the Poisson version in \refL{L1};
we omit the details. 
This completes the proof of Theorem~\ref{T1} for $p>1/2$.
The case $p<1/2$ is again the same by symmetry.
The proof when $p=1/2$ is similar, now using \eqref{axel}.
\end{proof}

\section{Proof of \refT{T3}}\label{Slc}

First, let us explain heuristically our estimate for $D_n(\alpha)$. 
By the Asymptotic Equipartition Property (\cf{} \cite{spa-book}) 
at level $k_n$ there are about  $n2^{-h k_n}$ strings with the same
prefix of length $k_n$ as a randomly chosen one,
where
$h$ is the entropy. That is, in the corresponding branch of the
$\ga$-LC trie, we have about 
$n2^{-h k_n} \approx n^{1-h/b}$ strings (or external nodes),
where for simplicity $b=\log(1/\sqrt{pq})$. In the next level, we shall have
about $n^{(1-h/b)^2}$ external nodes, and so on.
In particular, at level $D_n(\alpha)$ we have approximately
$$
                   n^{(1-h/b)^{D_n(\alpha)}}
$$
external nodes. Setting this $=\Theta(1)$ leads to our estimate
(\ref{eT3}) of Theorem~\ref{T3}.  
 
We now make this argument rigorous.
We construct an $\ga$-LC trie from $n$ random strings
$\xi_1,\dots,\xi_n$ and look at the depth $D_n(\alpha)$ of a
designated one of them. In principle, the designated string should be
chosen at random, but by symmetry, we can assume that it is the first
string $\xii$.
%As in \refS{Spo}, we note that the nodes in the (uncompressed) trie can
%be labelled by finite binary strings.

To construct the $\ga$-LC trie, we scan the strings 
$\xi_1,\dots,\xi_n$ in parallel one bit at a time, and build a trie
level by level.
As soon as the last level is filled less than $\ga$, we stop; 
we are now at level $F_n(\alpha)+1$, just past the \afillup{} level.
The trie above this level, \ie{} up to level $F_n(\alpha)$, is
compressed into one node, 
and we continue recursively with the strings attached to each node at
level $F_n(\ga)+1$ in the uncompressed trie, \ie{} the sets of strings
that begin with the same prefixes of length $F_n(\ga)+1$.

To find the depth $D_n(\ga)$ of the designated string $\xii$ in the
compressed trie, we may
ignore all branches not containing $\xii$; thus we let $Y_n$ be the
number of the $n$ strings that agree with $\xii$ for the first
$F_n(\ga)+1$ bits. Note that we have not yet inspected any later bits.
Hence, conditioned on $F_n(\ga)$ and $Y_n$, the remaining parts of
these $Y_n$ strings are again \iid{} random strings from the same
memoryless source, so we may argue by recursion. The depth $D_n(\ga)$
equals the number of recursions needed to reduce the number of strings
to 1.

We begin by analysing a single step in the recursion.
Let, for notational convenience, $\kk\=h/\log (1/\sqrt{pq})$.
Note that $0<\kk<1$.

\newcommand{\oo}{O\bigpar{n^{-\Theta(1)}}}

\begin{lemma}
  \label{LA}
Let $\eps>0$.
Then, with probability $1-\oo$,
\begin{equation}
  \label{c2}
1-\kk-\eps<\frac{\ln Y_n}{\ln n} < 1-\kk+\eps.
\end{equation}
\end{lemma}

We postpone the proof of \refL{LA}, and first use it to complete the
proof of \refT{T3}. We assume below that $n$ is large enough when
needed, and that $0<\eps<\min(\kk,1-\kk)/2$.

We iterate, and let $Z_j$ be the number of strings remaining after $j$
iterations; this is the number of strings that share the first $j$
levels with $\xii$ in the compressed trie. We have $Z_0=n$ and
$Z_1=Y_n$.
We stop the iteration when there are less than $\ln n$ strings
remaining; we thus let $\tau$ be the smallest integer such that
$Z_\tau<\ln n$. 
In each iteration before $\tau$, \eqref{c2} holds with error
probability 
$O\bigpar{(\ln n)^{-\Theta(1)}}
=O\bigpar{(\ln\ln n)^{-2}}$.
Hence, for any constant $B$,  we have \whp{}
for every $j\le \min(\tau,B\ln\ln n)$,
with $\kk_\pm=\kk\pm\eps\in(0,1)$,
\begin{equation*}
  1-\kk_+ < \frac{\ln Z_j}{\ln Z_{j-1}} < 1-\kk_-,
\end{equation*}
or equivalently 
\begin{equation}\label{c3}
  \ln(1-\kk_+) < \ln \ln Z_j - \ln\ln Z_{j-1} < \ln(1-\kk_-).
\end{equation}

If $\tau>\tau_+\=\ceil{\ln\ln n/\ln(1-\kk_-)\qi}$,
we find \whp{} from \eqref{c3} 
\begin{equation*}
  \ln\ln Z_{\tau_+} \le \ln\ln Z_0 + \tau_+ \ln(1-\kk_-) \le0,
\end{equation*}
so $Z_{\tau_+}\le e<\ln n$, which violates
$\tau>\tau_+$.
Hence, $\tau\le\tau_+$ \whp.

On the other hand, if 
$\tau<\tau_-\=\floor{(1-\eps)\ln\ln n/\ln(1-\kk_+)\qi}$,
then \whp{} by \eqref{c3}
\begin{equation*}
  \ln\ln Z_{\tau} \ge \ln\ln Z_0 + \tau_- \ln(1-\kk_+) \ge \eps \ln\ln n,
\end{equation*}
which contradicts $\ln\ln Z_\tau < \ln\ln\ln n$.

Consequently, \whp{} $\tau_-\le \tau\le \tau_+$; in other words,
we need $\frac{\ln\ln n}{-\ln(1-\kk)}\bigpar{1+O(\eps)}$ iterations to
reduce the number of strings to less than $\ln n$.

Iterating this result once, we see that \whp{} at most $O(\ln\ln\ln n)$
further iterations are needed to reduce the number to less than
$\ln\ln n$. Finally, the remaining depth then \whp{} is $O(\ln\ln\ln
n)$ even without compression.
Hence we see that \whp
\begin{equation*}
  D_n(\ga)
=
\frac{\ln\ln n}{-\ln(1-\kk)}\bigpar{1+O(\eps)}
+O(\ln\ln\ln n).
\end{equation*}
Since $\eps$ is arbitrary, \refT{T3} follows.

It remains to prove \refL{LA}. 
Let $W_k$ be the number of the strings $\xi_1,\dots,\xi_n$ that are
equal to $\xii$ for at least their first $k$ bits. The
$Y_n=W_{F_n(\ga)+1}$, and thus, for any $A>0$,
\begin{multline*}
\P\bigpar{|\log Y_n-(1-\kappa)\log n|\ge2\eps \log n}
\le
  \P\bigpar{|F_n(\ga)-\log_{1/\sqrt{pq}}n|\ge A \sqrt{\ln n}}
\\
+\hskip-1em
\sum_{|k-1-\log_{1/\sqrt{pq}}n|< A \sqrt{\ln n}}
\P\bigpar{|\log W_k-\log n+h\log_{1/\sqrt{pq}}n|\ge2\eps \log n}.
\end{multline*}
\refL{LA} thus follows from the following two
lemmas, using the observation that $0<1/\log(1/\sqrt{pq})<1/h$. 
%by summing \eqref{lc} over all $k$ in the range given by \refL{LB}.

The first lemma is a large deviation
estimate corresponding to \refC{CT1}.
\begin{lemma}\label{LB}
  For each $\ga\in(0,1)$, there exists a constant $A$ such that 
  \begin{equation*}
\P\bigpar{|F_n(\ga)-\log_{1/\sqrt{pq}}n|\ge A \sqrt{\ln n}}=O(1/n).
  \end{equation*}
\end{lemma}

\begin{proof}
  We begin with the poissonized version, with $\Po(\gl)$ strings as in
  \refS{Spo}.
Let 
$k_\pm=k_\pm(\gl)\=\floor{\log_{1/\sqrt{pq}}\gl\pm A\sqrt{\ln \gl}}$, 
and let $\gd$ 
be fixed with $0<\gd<\min(\ga,1-\ga)$.
Then, by \refL{L1}, if $A$ is large enough, 
$\E\bX_{k_-}>\ga+\gd$ and 
$\E\bX_{k_+}<\ga-\gd$ for all large $\gl$.
By a Chernoff bound, \eqref{bx} can be sharpened to 
\begin{equation*}
  \P\bigpar{|\bX_k-\E\bX_k|>\gd} =O\bigpar{e^{-\Theta(2^{k})}}
\end{equation*}
and thus
\begin{equation*}
  \begin{split}
  \P\bigpar{\tfla<k_-}
&\le \P\bigpar{\bX_{k_-}<\ga}
\le \P\bigpar{\bX_{k_-}-\E\bX_{k_-}<-\gd}
\\&
=O\Bigpar{e^{-\Theta(2^{k_-})}}
=O\Bigpar{e^{-\Theta(\gl^{O(1)})}}
=O(\gl^{-1}).	
  \end{split}
\end{equation*}
Similarly, $\P(\tfla>k_+)=O(\gl^{-1})$.

To depoissonize, let 
$\gl_\pm=n\pm n^{2/3}$ as in \refS{Sdepo}
and note that, again by a Chernoff estimate,
$\P\bigpar{\Po(\gl_-)\le n}=O(n^{-1})$ 
and
$\P\bigpar{\Po(\gl_+)\ge n}=O(n^{-1})$.
Thus, 
with probability $1-O(1/n)$, 
$$k_-(\gl_-)\le\tfxa{\gl_-}\le F_n(\ga) \le \tfxa{\gl_+}\le
k_+(\gl_+),
$$
and the result follows 
(if we increase $A$).
\end{proof}

\begin{lemma}
  \label{LC}
Lat $0<a<b<1/h$ and $\eps>0$. Then, uniformly for all $k$ with 
$a\log n\le k\le b\log n$,
\begin{equation}
  \label{lc}
\P\bigpar{|\log W_k-\log n+kh|>\eps \log n} = \oo.
\end{equation}
\end{lemma}

\begin{proof}
Let $\ones$  be the number of 1's in the first $k$ bits of $\xii$.
Given $\ones$, the distribution of $W_k-1$ is
$\Bi(n-1,p^{\ones}q^{k-\ones})$.

Since $p^pq^q=2^{-h}$, there exists $\gd>0$ such that if
$|\ones/k-p|\le\gd$, then
$2^{-h-\eps} \le p^{\ones/k}q^{1-\ones/k} \le 2^{-h+\eps}$, and thus
\begin{equation}\label{lc1}
  2^{-hk-\eps k} \le p^{\ones}q^{k-\ones} \le 2^{-hk+\eps k},
\qquad \text{when } |\ones/k-p|\le\gd.
\end{equation}
Noting that $hk\le bh\log n$ and $bh<1$, we see that, provided $\eps$
is small enough, $n 2^{-hk-\eps k} \ge n^{\eta}$ for some $\eta>0$, and
then \eqref{lc1} and a Chernoff estimate yields,
when $|\ones/k-p|\le\gd$,
\begin{equation*}
\P\bigpar{\tfrac12 n 2^{-hk-\eps k} \le W_k \le 2 n 2^{-hk-\eps k}
 \mid \ones}
= 1 - O\bigpar{e^{-\Theta(n^\eta)}}
=1-O\bigpar{n\qi},
%\qquad \text{when } |\ones/k-p|\le\gd.
\end{equation*}
and thus
\begin{equation}\label{lc2}
\P\bigpar{|\log W_k-\log n +hk| >\eps k +1
 \mid \ones}
=O\bigpar{n\qi},
\qquad \text{when } |\ones/k-p|\le\gd.
\end{equation}

Moreover, $\ones\sim\Bi(k,p)$, so by another Chernoff
estimate,
\begin{equation*}
  \P\bigpar{|\ones/k-p|>\gd} 
= O\bigpar{e^{-\Theta(k)}}
=\oo.
\end{equation*}
The result follows (possibly changing $\eps$) from this and \eqref{lc2}.
\end{proof}

\newcommand\vol{\textbf}

\end{document}